\documentclass[a4paper] {article}
\usepackage{graphicx}
\begin{document}

\large{\bf{Borexino}}\\

Lino Miramonti \textit{on behalf of the Borexino Collaboration}\\

\small{\textit{Department of Physics of University of Milano and INFN, via Celoria 16, I 20133 Milano, Italia}}

\begin{abstract}
Borexino is a massive calorimetric liquid scintillation detector whose installation has been
completed in the underground Gran Sasso Laboratory. The focus of the experiment is on
the direct and real time measurement of the flux of neutrinos produced in the $^{7}Be$
electron capture reaction in the Sun. Furthermore, recent studies about the reduction of
the $^{11}C$ background through suitable rejection techniques demonstrated the possibility to
open an interesting additional observation window in the energy region of the pep and CNO
solar neutrinos. Beyond the solar neutrino program, the detector will be also a powerful
observatory for antineutrinos from Supernovae, as well as for geoneutrinos, profiting from
a very low background from nuclear reactors.
\end{abstract}

\section{Goals}
The Borexino main goal is to detect neutrinos from the Sun via the scattering process
$e^{-}\nu_{x} \rightarrow e^{-}\nu_{x}$.
The neutrino signature is given by the scintillation light produced as the recoil
electron deposits its energy in the medium. The detector is designed to work with a very low energy 
threshold (250 keV), thus providing important information on the low energy portion of the solar neutrino spectrum. 
Recent studies proved the capability to reduce the $^{11}$C background through suitable rejection techniques and 
demonstrated the possibility to exploit the energy region of the pep and CNO solar neutrinos.

Borexino will also be able to observe antineutrinos coming from Supernovae and from the Earth 
(geoneutrinos); the detector is far away from nuclear reactors, which represent the most important
background to electron antineutrino detection.

\section{Status}
The installation of the detector has been completed in July 2004. Several campaigns of
data-taking with the empty detector have been performed in order to
check the overall performance of the apparatus, DAQ, electronics and software. 
Some runs were performed inserting a radiactive source into the detector (i.e. a vial containing
liquid scintillator loaded with Rn) to check the light collection efficiency of the detector and to
test the position reconstruction algorithm.
Currently (September 2006) the detector is being filled with ultra pure water and the filling with pseudocumene is foreseen starting from the end of this year. Borexino will be ready to take data by the spring of 2007.

\smallskip

\end{document}